\documentclass[acmlarge, authorversion=true]{acmart}

\pdfoutput=1
\makeatletter
\def\mdseries@tt{m}
\makeatother
\usepackage[draft=true]{minted}
\usepackage[minted, listings, breakable, skins]{tcolorbox}
\usepackage{amsmath}
\usepackage{amsfonts}
\usepackage{amsthm}
\usepackage{setspace}
\usepackage{mathtools}
\usepackage{subcaption}
\usepackage{suffix}
\RequirePackage{color}
\usepackage{adjustbox}
\include{math-commands}
\author{Brandon T. Willard}
\affiliation{\institution{University of Chicago} \city{Chicago} \state{IL} \country{USA}}
\email{bwillard@uchicago.edu}
\definecolor{bg}{rgb}{0.95,0.95,0.95}
\usepackage{natbib}
\usepackage{cleveref}
\allowdisplaybreaks
\setkeys{Gin}{keepaspectratio}
\graphicspath{{../../figures/}{../figures/}{./figures/}{./}}
\setminted{breaklines=true, breakanywhere=true, breakautoindent=true}
\usepackage{adjustbox}

\newlength{\maxtabfigwidth}
\newlength{\maxtabfigheight}
\usepackage{mleftright}
\usepackage{tcolorbox}
\tcbuselibrary{minted, listings, breakable, skins}
\date{2020-05-09}
\title{miniKanren as a Tool for Symbolic Computation in Python}
\hypersetup{
 pdfauthor={Brandon T. Willard},
 pdftitle={miniKanren as a Tool for Symbolic Computation in Python},
 pdfkeywords={},
 pdfsubject={},
 pdfcreator={Emacs 28.0.50 (Org mode 9.3.6)},
 pdflang={English}}
\begin{document}

\newtcblisting[auto counter,number within=section]{oxtcblisting}[1]{%
	frame hidden,
	listing only,
	listing engine=minted,
	minted options={linenos, numbersep=2mm},
	breakable,
	enhanced,
	title after break={\raggedleft\lstlistingname\ \thetcbcounter~ -- continued},
	listing remove caption=false,
	arc=0pt,
	outer arc=0pt,
	boxrule=0pt,
	coltitle=black,
	colbacktitle=white,
	center title,
	#1
}
\begin{abstract}
In this article, we give a brief overview of the current state and future
potential of symbolic computation within the Python statistical modeling and
machine learning community.  We detail the use of miniKanren
\citep{ByrdRelationalProgrammingminiKanren2009} as an underlying framework for
term rewriting and symbolic mathematics, as well as its ability to orchestrate
the use of existing Python libraries per
\citet{RocklinMathematicallyinformedlinear2013}.  We also discuss the
relevance and potential of relational programming for implementing more robust,
portable, domain-specific ``math-level'' optimizations--with a slight focus on
Bayesian modeling.  Finally, we describe the work going forward and raise some
questions regarding potential cross-overs between statistical modeling and
programming language theory.
\end{abstract}

\maketitle



\section{Introduction}
\label{sec:intro-symbolic-computation}
Throughout, we will focus on two categories of tools within the modern
machine learning, statistics, and data science world:  \textbf{Tensor Libraries} and
\textbf{Probabilistic Programming Languages} (PPLs).

For our purposes, it's sufficient to say that tensor libraries are the modern
wrappers for standard linear algebra operations--traditionally offered by BLAS
\citep{BLASBasicLinear2020} and LAPACK \citep{LAPACKLinearAlgebra2020}--with
extensions to handle general arrays, perform tensor operations, and
efficiently compute gradients--usually via automatic differentiation (AD).
These libraries are the main workhorses of deep learning (DL) libraries, and,
because of this strong association, tensor libraries often provide
deep learning-specific functionality (e.g. tools for constructing
DL models, common activation functions, etc.)

Probabilistic programming languages are domain-specific languages that aid in
the specification of statistical models and the application of estimation
methods on said models.  Often, PPLs will reflect the formal, probability
theory-based language used to specify statistical models, but that connection
with probability theory tends to serve primarily in an interface role.
PPLs also implement related elements--like random variables and statistical
distributions--and, in some cases, they provide limited support for the laws and
identities of probability theory (e.g. addition and multiplication of random
variables).

Nowadays, PPLs are increasingly backed by tensor libraries, so the two subjects
are connected in this way.

There is an appreciable amount of symbolic computation behind the standard Deep
Learning libraries of today, like Theano \citep{bergstra_theano:_2010},
TensorFlow \citep{TensorFlow2020}, and PyTorch \citep{PyTorch2020}.

At the very least, these libraries provide classes that represent tensor algebra
in graphical form and functions that manipulate graphs.  Furthermore, these
graphs are used to compute derivatives--via Automatic Differentiation (AD),
parallelize or vectorize operations, and, in limited cases, explicitly perform
algebraic simplifications.

Regardless of whether or not you're in a camp that says AD falls well within
standard symbolic mathematics, these libraries are undoubtedly performing
symbolic calculations, and most of them are on a path toward even more symbolic
computation and outright symbolic mathematics.

Theano's optimizations are perhaps the best example of more open-ended symbolic
computation within modern tensor libraries.  It provides an entire subsystem for
incorporating custom ``graph optimizations'' that implements pattern matching,
term substitution, utilizes common sub-expression elimination (CSE), offers
multiple graph traversal strategies and fixed-point operators, etc.  The
optimization system is used to perform graph canonicalization and specialization
through a number of standard matrix algebra identities, and, with these, it is
able to avoid numerous unnecessary numeric calculations and increase the overall
``flexibility'' of function/model specification.

Unfortunately, most other tensor libraries do not offer as much in the way of
user-level Python manipulation of graphs.  In the case of TensorFlow, an
internal graph canonicalization and optimization library Grappler
\citep{LarsenTensorFlowGraphOptimizations2019a} is being actively developed.
At the moment--its core work is done exclusively in C++ and the potential
for robust user-defined optimizations isn't clear.

Outside of the aforementioned tensor libraries, other projects approach similar
symbolic-oriented operations--like AD--through function tracing
\citep{JAX2020} and Python AST parsing \citep{tangent2020}.  As with most
tensor libraries, these projects aren't particularly focused on supporting
generic graph manipulation or symbolic mathematics.  Nevertheless, they
demonstrate another burgeoning entry-point to general symbolic computation.

At this point in time, there is a fairly well established set of modern
machine learning and statistical modeling libraries that are all fundamentally
built upon the basics of symbolic graphs representing
tensor algebra operations.  They all vary in the degree to which they
implement symbolic mathematics or programmatically encode the underlying
high-level math.  Regardless, the case for more advanced symbolic
computation and mathematics is slowly being made by multiple influential
projects, and there's already enough reason and prior work to start assessing
the directions this could go, and how we might get the most out of it.

Ideally, symbolic work within the machine learning and statistical modeling
community would progress by developing constructively on top of well established
projects, so that quality code can be reused, existing expertise can be
leveraged, and community involvement from domain experts is easily incorporated.

Within the Python community, \citet{RocklinMathematicallyinformedlinear2013}
states the same sentiments with regards to the intelligent use of optimized
linear algebra functions and the basics of term rewriting.  This work follows
the same principles, but focuses on the areas of statistical modeling and, more
specifically, Bayesian modeling.  That is to say, we seek flexible,
compartmentalized and open systems that are easily integrated with established
libraries and map well to their high-level abstractions.  Ideally, we could have
all this without much context switching--in other words, the need to operate in
more than one programming languages or between programming languages.

\section{What we want to do in Statistical Modeling}
\label{sec:want-to-do}
We believe that the statistical modeling and machine learning communities need a
framework within which they can develop their own high-level, symbolic
"optimizations" specific to the domains of statistical modeling and machine
learning.  Ideally, such a framework would have the properties outlined in
Section \ref{sec:intro-symbolic-computation} and build upon the already well
established Python machine learning and statistics ecosystem, instead of
attempting to outright reinvent it or rewrite its staple offerings.

These symbolic optimizations should be usable internally by new and existing
libraries to drive advanced automations.  As well, they should be usable in an
\textbf{interactive} way by researchers, where the researchers themselves can dynamically
add new theorems and immediately use the resulting automations for
high-level testing, experimentation, and concrete applications.

Statistical modeling is surprisingly amenable to symbolic methods
\citep{CaretteCasestudiesmodel2008,CaretteSimplifyingProbabilisticPrograms2016}--and especially when one
restricts the context to specific practices like Bayesian modeling
\citep{ShanExactBayesianInference2017}, where there exist fundamental
relations--such as "conjugacy" \citep[Chapter 3.3]{robert_bayesian_2007}--that are
easily automated with simple pattern matching.

Another example is Rao-Blackwellization.  Rao-Blackwellization is derived from
the Rao-Blackwell Theorem
\citep{CasellaRaoBlackwellisationsamplingschemes1996}, which
states--roughly--that analytically computed conditional expectations outperform
their un-integrated counterparts.  In other words, if you can get a closed-form
answer to an integral, instead of estimating the integral with samples, you
should use the closed-form answer.

It applies in a rather general sense to numerous Markov Chain Monte Carlo
methods, but its automation is something that falls well outside of most
libraries, arguably due to its symbolic computation requirements.

The Rao-Blackwellizations appearing in published material are often driven by
simple high-level identities--identities which we can alternatively classify as
\textbf{relations}.  Some of those relations reflect basic theorems in probability
theory and statistics, like the following normal (or Gaussian) random variable
identity--expressed as a rule:
\begin{equation}
  \label{eq:sum-of-normals}
  \begin{aligned}
    \left(\text{sum-of-normals}\right) & \quad
    \frac{
      X \sim \operatorname{N}\mleft(\mu_x, \sigma_x^2\mright), \quad
      Y \sim \operatorname{N}\mleft(\mu_y, \sigma_y^2\mright), \quad
      X + Y = Z
    }{
      Z \sim \operatorname{N}\mleft(\mu_x + \mu_y, \sigma_x^2 + \sigma_y^2\mright)
    }
  \end{aligned}
\end{equation}

There are numerous examples like \eqref{eq:sum-of-normals}, and they all take the
form of relations.  Hiding behind these relations are the closed-form integrals
that would otherwise be painstaking to compute directly with symbolic algebra.

As a matter of fact, there are at least two ways to frame identities like these:
in terms of random variables, and in terms of their corresponding distribution
functions.  This means that one can turn theorems like Rao-Blackwellization
into an integration problem.  Unfortunately, term graphs produced by the
distribution-based approach can be much more complex than the corresponding random
variable graphs.  Our focus will be on the latter approach, since it
offers more opportunities to solve equivalent integration problems using only
collections of simple random variable identities.

This general idea has analogs in the approaches used by modern symbolic
integration systems themselves.  When such systems employ Fox H and Meijer G
functions \citep{peasgood_method_2009,roach_meijer_1997}, they are
effectively using only a few simple algebraic convolution identities applied to
broad classes of hypergeometric functions--many of which can be encoded by
simple look-up tables.

Since those systems are intended to reach a much greater number of functions and
constraints, they are necessarily more complex; however, a majority of the work
being done by statistical modeling deals with a comparatively smaller set of
standard distributions, so major improvements can be made without invoking the
complexity of symbolic integration systems.

Currently, statistical modeling systems do not directly support these types of
``knowledge'' additions, nor do they attempt to systematically employ these
well-known and far-reaching theorems--like Rao-Blackwellization.  Instead, this
kind of work is still restricted to the user-level, where it is performed by
hand and used as input to such systems.  At the present, the best systems simply
provide broadly useful identities and theorems as advice in their manuals and
message boards.

In particular, Stan \citep{standevelopmentteam_stan_2014} is known for
having a well written manual that details user-level manipulations to account for
common sampling issues arising due to poorly specified models
\citep{GelmanTransformingparameterssimple2019}.  Note that the description
"poorly specified" is conditional on the given estimation approach.

Among the advice given in Stan's manual is the classic pathological "funnel"
model of \citet{neal_slice_2003}.  This model can be reparameterized using the
following rule between a standard Gaussian random variable and its
affine transform:
\begin{equation}
  \label{eq:normal-affine-trans}
  \begin{aligned}
    \left( \text{normal-affine-transform} \right) &\quad
    \frac{
      Y \sim \operatorname{N}\mleft( 0, 1 \mright), \quad
      X = \mu + \sigma Y
    }{
      X \sim \operatorname{N}\mleft( \mu, \sigma^2 \mright)
    }
  \end{aligned}
\end{equation}

Under \eqref{eq:normal-affine-trans}, terms in the funnel model can be expanded
resulting in an equivalent model that exhibits much better sampling properties.

The work we detail here is motivated by the desire to see relations like these
used within statistical modeling systems, so that model specification is more
flexible and less brittle with respect to the exact specification of a model.

Systems like Stan and the Python-based PyMC
\citep{SalvatierProbabilisticprogrammingPython2016} are PPLs, so their role as
programming languages is clear, and--in line with most programming
languages--compiler optimizations can be used to improve performance
and expand the expressive potential of a language's syntax.

Projects like Stan and PyMC rely almost exclusively on AD and have more or less
superseded older projects based on different, non-gradient-based generalized
methodologies, like BUGS \citep{LunnBUGSprojectEvolution2009}.  BUGS used some
of the domain-specific identities implied here to construct a surprisingly robust
expert system that could automatically construct a sampler for a
given model.  We would like to make such systems easier to produce and extend,
and we would like to see them built on top of tensor libraries, so that
the AD-driven methods of modern PPLs can be used in tandem.

One noteworthy example is PyMC's internal logic for determining an appropriate
sampler.  This logic could benefit from an easily extensible, expert-like system
that matches models to samplers.  Just like the optimization system in Theano,
the graph of a PyMC model can be manipulated to produce a more suitable, yet
equivalent, model for a given sampler, or--conversely--produce a customized
sampler for a given model.  In extreme cases, the posterior distribution
ultimately estimated by PyMC could be returned in closed-form.  A small example
of this is given in Section \ref{sec:symbolic-pymc}.

Otherwise, there are entire classes of efficient, model-specific samplers that
are currently out of these PPLs' reach, and the addition of some straight-forward and
flexible term rewriting capabilities would make them immediately available.
Some examples involve Gibbs samplers, scale mixture representations for sparsity priors
\citep{BhadraDefaultBayesiananalysis2016}, non-Gaussian models
\citep{polson_bayesian_2013}, and parameter expansions
\citep{scott_parameter_2010}.

As a proof of concept using Theano's existing optimization system, automatic
simplification of random variables was demonstrated in
\citet{WillardRoleSymbolicComputation2017}.  While it is more than possible to
extend the same approach into auto-conjugation and related statistical
optimizations, the scalability and means of specifying new optimizations within
Theano wasn't promising.

One important concern involves the need to use identities in more than one
direction.  For instance, one direction of the identity underlying
\eqref{eq:normal-affine-trans} is useful for computational reasons (e.g. the
funnel problems) and the other direction helps one determine the distribution
type of sub-term (i.e. given \(Y \sim \operatorname{N}\mleft( 0, 1 \mright)\)
we can derive distribution of \(X\)) in a larger model.  The latter information
might be needed by a system that constructs custom samplers, or to reformulate a
model so that it can be used by a given sampling routine.

This otherwise natural use of identities isn't covered well by modern
programming frameworks, and that's where logic and relational programming
becomes a real consideration.

Considerations like these also lead quickly into the domain of term rewriting
\citep{BaaderTermrewritingall1999}.  Graph normalization/canonicalization,
rewrite rule completion \citep{Huetcompleteproofcorrectness1981}, and general
equational reasoning are all ground-level subjects in the features we've described.

With this in mind, it's likely that our objectives won't often lead to the
classical term rewriting niceties, like easily determined normal forms and term
orderings with strong guarantees.  Given our desire for an interactive system in
which to perform ad hoc additions and experimentation, it seems even less
likely.  Even so, when such niceties are available, we would at least like a
suitable framework in which to derive and apply them.  Furthermore, if it's ever
possible to produce \emph{any} results from a less-than-perfect set of identities,
then we would like a framework that facilitates that, too.

Overall, we seek a middle ground that provides an approachable, powerful, yet
light-weight framework for creating and orchestrating domain-specific relations.

As well, we would like this framework to promote joint development between
experts in statistics, machine learning, term rewriting, type theory, code
synthesis, and related areas.  We believe miniKanren could serve an important
role within this intersection of requirements.

\section{Where miniKanren fits in}
\label{sec:org4664062}

Computer science researchers have been--and continue to--actively pursue topics
in symbolic computation specifically within the area of statistical modeling
\citep{Waliahighlevelinferencealgorithms2018,SatoFormalverificationhigherorder2018,ShanExactBayesianInference2017}.
While very in-line with the automations described here, much of this work takes
the form of entirely new languages or very broad theoretical work that doesn't
always lend itself to more immediate input from experts in the areas of
statistical modeling methods.

In other cases, the limitations involve the degree of specialization, where
exclusive focus is often on neural network-specific DSLs and frameworks, or
only certain types of optimizations \citep{WeiDLVMmoderncompiler2017,VasilacheTensorComprehensionsFrameworkAgnostic2018}.

Regarding Python and statistical modeling, the recent automatic conjugation and
rescaling examples of \citet{HoffmanAutoconjRecognizingExploiting2018,GorinovaAutomaticReparameterisationProbabilistic2018} are perhaps the most germane;
however, their approach relies entirely on an existing pattern-matching and
rewrite system \citep{RadulRules2020} that is non-relational and uses the
Python stack for backtracking.  As we've stated earlier, the use of relations
has important conceptual and implementation advantages (e.g. the concepts being
implemented are fundamentally relational, and the inherent code reuse arising
from "bidirectional" applications of identities).

As well, use of the Python stack for backtracking puts severe limitations on the
size of graphs manageable by such a system.  Python
throws \mintinline{python}{RecursionErrors} when the stack reaches a fixed
size, and term graphs representing real models are by no means small, so
unification alone is liable to cause irreconcilable errors.  Our implementation
of unification in Python \citep{Willardlogicalunification2020} demonstrates
this exact problem in its unit tests, and, as a result, the library uses a
coroutine-based trampoline to avoid excessive use of the Python stack.

Simply increasing the recursion limit (e.g.
via \mintinline{python}{sys.setrecursionlimit}) is--at best--a single-case
solution, and it's often safer to pursue a more "pythonic" rewrite (i.e. loop or
list comprehension-based approach).  Also, Python currently lacks even the most basic
forms of tail recursion elimination--and it's very unlikely to appear in later
versions \citep{RossumNeopythonicTailRecursion2009a}.

Furthermore, \citet{HoffmanAutoconjRecognizingExploiting2018} doesn't provide
clear examples of how rewrite rules are specified in their proposed system, so
it's difficult to assess exactly how expressive their DSL is, or even how well it
works within Python and its standard collection types.

This is where miniKanren comes in.  It serves as a minimal, lightweight
relational DSL that orchestrates unification and reification and operates
exclusively within an existing host language.  Furthermore, its core
functionality is succinctly described in a single page of code, which helps make
its inner workings very transparent to the interested developer
\citep{HemannmKanrenminimalfunctional2013}.

In contrast with other unification-driven systems, its "internal" mechanics maintain direct
connections with multiple high-level theoretical concepts (e.g. unification,
complete search, relational programming, CPS, etc.), so, for--instance--its use
as a type theory prototyping language automatically provides exciting connections to
both basic and cutting-edge symbolic computation (e.g. automatic theorem proving
\citep{NearalphaleanTAPDeclarativeTheorem2008}).  As a matter of fact, the use
of typing rules to describe the automation of high-level inference algorithms in
\citet{Waliahighlevelinferencealgorithms2018} is a direct example of how
elements of type theory can be used in high-level statistical model optimization, and
miniKanren can serve as a bridge to fast implementations.

miniKanren inherently provides a degree of high-level portability and low-level
flexibility.  Relations can be built on top of other relations, and, in these
cases, the goals that implement such composite relations in miniKanren are often
easily ported to miniKanrens in other host languages.  miniKanren doesn't
enforce a formal, host-language independent semantics, yet it still lends well
to this kind of portability.  This lack of formal semantics also makes it easier
to address performance and domain specific issues in multiple ways--like
the \mintinline{python}{RecursionError}s described above.
With these properties, miniKanren has exactly the type of generality and
flexibility to serve as a basis for more fluid collaboration--in symbolic
computation--between independent communities of computer scientists and
statisticians.

In the following section, we will illustrate many of these points using our
Python implementation of miniKanren.

\section{Symbolic Computation in Python Driven by miniKanren}
\label{sec:orgbcb62f7}

In the following sections we detail our implementation of miniKanren
\citep{Willardpythologicalkanren2020}, operating under the PyPi
name \mintinline{python}{miniKanren} and Python package
name \mintinline{python}{kanren}.  We also describe its ecosystem of
complementary packages \mintinline{python}{etuples}, \mintinline{python}{cons},
and \mintinline{python}{symbolic-pymc}.

\mintinline{python}{kanren} is a fork that tries to maintain syntactic
parity with its predecessor \mintinline{python}{logpy}
\citep{RocklinlogpyLogicProgramming2018}, but now deviates significantly in terms
of core mechanics and offerings.  The most important difference is in the
relational status of \mintinline{python}{logpy}'s goals; most were not truly
relational.  This was largely due to the use of
an exception-based goal reordering system, which served as the exclusive
means of handling missing \mintinline{scheme}{cons}-based capabilities and minimalistic constraints.
Basically, one could attempt to develop entirely in standard Python at the goal
constructor level, and throw special \mintinline{python}{EarlyGoalError}
exceptions when goal constructor arguments were not sufficiently ground for
a given task.  The exception would cause the goals to be reordered until
an ordering satisfied these goals' groundedness requirements.

This exception-based approach was combined with a lightweight tuple-based,
Lisp-like expression evaluator that operated in tandem with the goal reordering.  Both
of these components were built directly into the core stream processing
functions and introduced additional complexity and challenges, but, most of all,
they made it much easier to construct non-relational goals and imposed new,
non-miniKanren semantics that increased the barrier to entry beyond a simple
understanding of core Python and miniKanren.

Our implementation of miniKanren's core mechanics does not operate on Lisp-like
idioms, yet it maintains an operational similarity to the Scheme implementations.
Additionally, it provides a straight-forward object-oriented framework for
implementing truly relational constraints.  These points will be covered in more
detail in the following sub-sections.

First, we must note that both Python implementations share the same small, but
noteworthy, deviations from the standard Scheme-based miniKanrens.
Specifically, the basic miniKanren states are implemented using Python's
built-in \mintinline{python}{dict} type, the \(\equiv\) goal is represented
by the function \mintinline{python}{eq}, there is no \texttt{fresh}--instead, fresh
logic variables are constructed explicitly using the
function \mintinline{python}{var}--and the functionality of \texttt{bind} and
\texttt{mplus} are represented by the logical ``and'' and ``or''
functions \mintinline{python}{lall} and \mintinline{python}{lany} and
are essentially the \texttt{conj} and \texttt{disj} stream functions of
\citet{HemannmKanrenminimalfunctional2013}.

\subsection{Goals as Generators}
\label{sec:org5e1fe9d}
Our implementation of miniKanren represents goals using
Python's built-in generators \citep{GeneratorsPythonWiki2020,PythonSoftwareFoundationExpressionsPythondocumentation2020}.

Listing \ref{low-level-relations} illustrates the general form of a miniKanren goal
and some of the idioms available to them.

\begin{oxtcblisting}{minted language=python, title={\lstlistingname\ \thetcbcounter: {Example idioms for generator-based goals in Python.}},label type=listing, label={low-level-relations},nofloat}
def relationo(*args):
    """Construct a goal for this relation."""

    def relationo_goal(S):
        """Generate states for the relation `relationo`.

        I.e. this is the goal that's generated.

        Parameters
        ----------
        S: Mapping
            The miniKanren state (e.g. unification mappings/`dict`).

        Yields
        ------
        miniKanren states.

        """
        nonlocal args

        args_rf = reify(args, S)

        x = 1
        for a in args_rf:
            S_new = S.copy()

            if isvar(a) or x > 3:
                S_new[a] = x

            z = yield S_new  #

            if not z:
                x += 1

        if some_condition:
            yield S  #
        else:
            return

        a_lv = var()
        yield from lall(conso(1, a_lv, args), eq(a_lv, [2, 3]))  #

        yield from relationo(*new_args)  #

    return relationo_goal

\end{oxtcblisting}

Simply put, a goal is responsible for either explicitly generating its goal
stream (e.g. Line 30 and 36) or deferring to
other goals and/or goal combinations via the
stream manipulation functions \mintinline{python}{lall}
and \mintinline{python}{lany} (e.g. Line 41).

The idioms described in Listing \ref{low-level-relations} are realized in a number of
low-level goal implementations within \mintinline{python}{kanren}.  One good
example is the \mintinline{python}{permuteo} goal, which relates an ordered
collection to its permutations.  Within \mintinline{python}{permuteo},
low-level Python steps are taken in order to efficiently compute differences of
hashable collections when the arguments are ground, and, when one argument
is unground, Python's built-in permutation
generator \mintinline{python}{itertools.permutations} is used to efficiently
generate unification arguments for the unground term.  This strictly
Python-based low-level implementation of a goal is both completely relational
and considerably more scalable than an implementation built on the basic
miniKanren relations and amounting to Bogosort
\citep{KiselyovBacktrackinginterleavingterminating2005}.

Within ordinary goals like \mintinline{python}{relationo_goal} one is able
to leverage the naturally delayed nature of Python's generators and seamlessly
define recursive goals (e.g. Line 43) by calls to the outer goal
constructor \mintinline{python}{relationo}, and, in the case of goal
constructors that do not define their own low-level goals, recursion is
facilitated by the \(\eta\)-delay function, \mintinline{python}{Zzz}, of
\citet{HemannmKanrenminimalfunctional2013}.

This approach also makes it possible for goals to more easily control the type
and order of the results it streams.  The loop around Line 30
in Listing \ref{low-level-relations}, demonstrates how a goal can easily keep and
manage its state--e.g. the variable \mintinline{python}{x}--and use it to affect
the goal stream it produces.

Also, using Python's coroutine capabilities, Line 30 shows
how it's possible to send results back to a goal when the process evaluating the
stream uses \mintinline{python}{generator.send} \citep{PEP342Coroutines2020}.
In this case, a goal could be given ``upstream'' information.

Also, using Python's \mintinline{python}{__length_hint__}
\citep{PEP424method2020} spec, goals and stream manipulation functions could be
told when a stream is empty or simply make decisions based on partial information about a
stream's size.  Such information could help determine efficient orderings
within \mintinline{python}{lall} conjunctions and
between \mintinline{scheme}{conde} branches, by--say--allowing these
operators to choose finite streams over potentially infinite ones in certain cases.

Overall, the resulting simplicity of this approach is an example of how well
miniKanren's underlying mechanics can be adapted to host languages in which the
standard list-based approach isn't as natural or efficient as it is in Scheme.

\subsection{Constraints}
\label{sec:org7ebb51a}
Our Python implementation follows the approach of
\citet{Hemannframeworkextendingmicrokanren2017} to implement a minimal
constraint system in miniKanren.  Listing \ref{constraints-example} provides some
simple illustrations of the standard disequality
(named \mintinline{python}{neq} here) and type constraints--the latter using
Python's \mintinline{python}{isinstance} naming scheme.

\begin{oxtcblisting}{minted language=python, title={\lstlistingname\ \thetcbcounter: {Basic constraint goals example.}},label type=listing, label={constraints-example},nofloat}
>>> from kanren.constraints import neq, isinstanceo

>>> run(0, x,
...     neq(x, 1),  # Not "equal" to 1
...     neq(x, 3),  # Not "equal" to 3
...     membero(x, (1, 2, 3)))
(2,)

>>> from numbers import Integral
>>> run(0, x,
...     isinstanceo(x, Integral),  # `x` must be of type `Integral`
...     membero(x, (1.1, 2, 3.2, 4)))
(2, 4)

\end{oxtcblisting}

When constraints are used, the state type--normally an ordinary
Python \mintinline{python}{dict}--is replaced with a new
type: \mintinline{python}{ConstrainedState}.  This type is a subclass of the
interface \mintinline{python}{Mapping}, so it behaves effectively the same
as a standard \mintinline{python}{dict}.  The main functionality provided
by \mintinline{python}{ConstrainedState} is constraint store tracking and
validation.

Constraint store validation occurs after each successful unification
involving a \mintinline{python}{ConstrainedState}.  Adding new constraints
involves constructing a custom constraint store class and a goal that assigns
the constraint and adds or updates the associated store in a miniKanren state.
For convenience, there is an abstract \mintinline{python}{PredicateStore}
type that simplifies the construction of predicate-based constraints.

Given that constraint checking is tied directly to
the \mintinline{python}{Mapping} interface, one can dispatch on key
addition, deletion, and updating in order to implement more efficient constraint
store management and validation.

\subsection{\mintinline{scheme}{cons}}
\label{sec:org64beb48}
One of the main challenges involved in implementing miniKanren in some host
languages is the lack of immediate support for important Scheme/Lisp-like
elements.  The most notable for Python being \mintinline{scheme}{cons}.

\mintinline{scheme}{cons} support is important for maintaining certain forms
of simplicity and expressiveness in term rewriting.  For instance, while
``pattern matching''--or unification--alone can be rather straight-forward to
implement, and more than a couple of the software systems mentioned here have
introduced basic pattern matching, they all tend to lack the expressive
simplicity afforded by list-based terms and proper \mintinline{scheme}{cons}
semantics.

A good example is Theano's unification system
\citep{GraphoptimizationTheano2020}; although it does provide a tuple-based
interface for defining forms to match and replace, and it supports logic variables
within said forms, it doesn't provide a means of expressing
a \mintinline{scheme}{cons} pair.  As a result, attempting to construct a
pattern that matches a specific operator (or \mintinline{scheme}{car}) and
an unspecified number/type of arguments (or \mintinline{scheme}{cdr})--or
vice versa--falls outside of the system's reach.

Our Python implementation of miniKanren preserves nearly all the same algebraic
datatype semantics of Lisp's \mintinline{scheme}{cons} by way of
our \mintinline{python}{cons} package
\citep{Willardpythologicalpythoncons2020}.  The \mintinline{python}{cons}
package provides a minimal \mintinline{python}{ConsType} class, along with a
set of easily extensible generic functions for \mintinline{python}{car}
and \mintinline{python}{cdr}.

As Listing \ref{cons-cdr-unify-example} demonstrates,
with \mintinline{python}{cons} we're able to succinctly express the
aforementioned ``pattern'' for all the built-in ordered collection types, and
reify accordingly.
\begin{oxtcblisting}{minted language=python, title={\lstlistingname\ \thetcbcounter: {\mintinline{scheme}{cons} pair unification and reification using Python's built-in lists.}},label type=listing, label={cons-cdr-unify-example},nofloat}
>>> from collections import OrderedDict
>>> from cons import cons
>>> from unification import unify, reify, var

>>> unify([1, 2], cons(var('car'), var('cdr')), {})
{~car: 1, ~cdr: [2]}

>>> unify((1, 2, 3), cons(var('car'), var('cdr')), {})
{~car: 1, ~cdr: (2, 3)}

>>> unify(OrderedDict([('a', 1), ('b', 2)]), cons(var('car'), var('cdr')), {})
{~car: ('a', 1), ~cdr: [('b', 2)]}

>>> reify(cons(1, var('cdr')), {var('cdr'): [2, 3]})
[1, 2, 3]

>>> reify(cons(1, var('cdr')), {var('cdr'): (2, 3)})
(1, 2, 3)

\end{oxtcblisting}

Later, in Listing \ref{math-constrained-expand-reduce}, we provide another example
of how a \mintinline{scheme}{cons}-compliant unification system makes
non-trivial patterns easier to express.

\subsection{S-Expressions}
\label{sec:org28d2280}

Since Python doesn't already provide a programmatically convenient form of
expressions or terms, we've constructed a
simple
\mintinline{python}{ExpressionTuple} class--or \mintinline{python}{etuple}
for short--that extends the built-in \mintinline{python}{tuple} with
the ability to evaluate itself, cache the results, and maintain the cached
results between non-modifying reconstructions and re-evaluations of the
same \mintinline{python}{etuple}.

Python does provide AST objects that fully represent its built-in expressions,
but they are cumbersome to work with and do not provide much of the desired
functionality for term rewriting (e.g. access to nested elements is too
indirect, their construction and use involves irrelevant meta information, etc.)
See \citet{WillardReadableStringsRelational2018b} for examples of term
rewriting using Python AST objects and miniKanren.

As we demonstrate in a later example (i.e. Listing
\ref{math-reduceo}), \mintinline{python}{etuples} are an extremely convenient way
to leverage a target library's user-level functions (e.g. TensorFlow's matrix
multiplication function) without having to manually construct fully reifiable
term graphs--many of which require detailed information that may not be
available at the time of a goal's evaluation.

\begin{oxtcblisting}{minted language=python, title={\lstlistingname\ \thetcbcounter: {Constructing a simple \mintinline{python}{etuple}.}},label type=listing, label={etuple-examples},nofloat}
>>> from operator import add
>>> from etuples import etuple, etuplize

>>> et = etuple(add, 1, 2)
>>> et
ExpressionTuple((<built-in function add>, 1, 2))

\end{oxtcblisting}

\mintinline{python}{etuples} can be indexed--and generally treated--like
immutable \mintinline{python}{tuple}s:
\begin{oxtcblisting}{minted language=python, title={\lstlistingname\ \thetcbcounter: {\mintinline{python}{etuple} indexing example.}},label type=listing, label={etuple-index},nofloat}
>>> et[0:2]
ExpressionTuple((<built-in function add>, 1))

\end{oxtcblisting}

Evaluation is available through a simple cached property:
\begin{oxtcblisting}{minted language=python, title={\lstlistingname\ \thetcbcounter: {\mintinline{python}{etuple} evaluation example.}},label type=listing, label={etuple-eval},nofloat}
>>> et.eval_obj
3

\end{oxtcblisting}

Furthermore, it is easy to specify conversions to and
from \mintinline{python}{etuple}s for arbitrary types.

Listing \ref{etuple-custom-class} constructs two custom
classes, \mintinline{python}{Node} and \mintinline{python}{Operator},
and specifies the \mintinline{scheme}{car} and \mintinline{scheme}{cdr}
for the \mintinline{python}{Node} type via the generic functions
\citep{Rocklinmultipledispatch2019}
\mintinline{python}{rands} \mintinline{python}{rator}, respectively.
An \mintinline{python}{apply} dispatch is also specified, which represents a
combination of \mintinline{scheme}{cons} (via the aforementioned \mintinline{python}{cons}
library) and an S-expression evaluation.

\begin{oxtcblisting}{minted language=python, title={\lstlistingname\ \thetcbcounter: {Adding \mintinline{python}{etuple} support to a standard Python class.}},label type=listing, label={etuple-custom-class},nofloat}
from collections.abc import Sequence

from etuples import rator, rands, apply
from etuples.core import ExpressionTuple


class Node:
    def __init__(self, rator, rands):
        self.rator, self.rands = rator, rands

    def __eq__(self, other):
        return self.rator == other.rator and self.rands == other.rands


class Operator:
    def __init__(self, op_name):
        self.op_name = op_name

    def __call__(self, *args):
        return Node(Operator(self.op_name), args)

    def __repr__(self):
        return self.op_name

    def __eq__(self, other):
        return self.op_name == other.op_name


rands.add((Node,), lambda x: x.rands)
rator.add((Node,), lambda x: x.rator)


@apply.register(Operator, (Sequence, ExpressionTuple))
def apply_Operator(rator, rands):
    return Node(rator, rands)

\end{oxtcblisting}

With the specification of \mintinline{python}{rands}, \mintinline{python}{rator},
and \mintinline{python}{apply} for \mintinline{python}{Node} types, it is now
possible to convert \mintinline{python}{Node} objects to \mintinline{python}{etuples}
using the \mintinline{python}{etuplize} function.  Listing \ref{etuple-custom-etuplize}
demonstrates this process and shows how the underlying object is preserved through
conversion and evaluation.

\begin{oxtcblisting}{minted language=python, title={\lstlistingname\ \thetcbcounter: {Converting a supported class instance into an \mintinline{python}{etuple}.}},label type=listing, label={etuple-custom-etuplize},nofloat}
>>> mul_op, add_op = Operator("*"), Operator("+")
>>> mul_node = Node(mul_op, [1, 2])
>>> add_node = Node(add_op, [mul_node, 3])
>>> et = etuplize(add_node)

>>> pprint(et)
e(+, e(*, 1, 2), 3)

>>> et.eval_obj is add_node
True

\end{oxtcblisting}

\subsection{Relations for Term Rewriting}
\label{sec:orgcdb8eae}
Our Python implementation of miniKanren is motivated by the need to cover some of the
symbolic computation objectives laid out here, so, in response, it provides relations
that are specific to those needs.

The most important set of relations involve graph traversal and
manipulation.  \mintinline{python}{symbolic-pymc} provides ``meta'' relations
for applying goals to arbitrary ``walkable'' structures (i.e. collections that fully
support \mintinline{python}{cons} semantics via \mintinline{python}{car}
and \mintinline{python}{cdr}).

Listing \ref{math-reduceo} constructs an example goal that represents two simple
mathematical identities: i.e. \(x + x = 2 x\) and \(\log \exp x = x\).

\begin{oxtcblisting}{minted language=python, title={\lstlistingname\ \thetcbcounter: {An example goal that implements some basic mathematical relations.}},label type=listing, label={math-reduceo},nofloat}
def single_math_reduceo(expanded_term, reduced_term):
    """Construct a goal for some simple math reductions."""
    # Create a logic variable to represent our variable term "x"
    x_lv = var()
    return lall(
        # Apply an `isinstance` constraint on the logic variable
        isinstanceo(x_lv, Real),
        isinstanceo(x_lv, ExpressionTuple),
        conde(
            # add(x, x) == mul(2, x)
            [eq(expanded_term, etuple(add, x_lv, x_lv)),
             eq(reduced_term, etuple(mul, 2, x_lv))],
            # log(exp(x)) == x
            [eq(expanded_term, etuple(log, etuple(exp, x_lv))),
             eq(reduced_term, x_lv)]),
    )

\end{oxtcblisting}

We can combine the goal in Listing \ref{math-reduceo} with the ``meta''
goal, \mintinline{python}{reduceo}, which applies a goal recursively until a
fixed-point is reached--assuming the relevant goal is a reduction, of course.
(The meta goal should really be named \mintinline{python}{fixedpointo}.)
Additionally, we create another partial function that sets some default arguments
for the \mintinline{python}{walko} meta goal.

\begin{oxtcblisting}{minted language=python, title={\lstlistingname\ \thetcbcounter: {Partial functions for a fixed-point calculation and graph walking.}},label type=listing, label={math-partials},nofloat}
math_reduceo = partial(reduceo, single_math_reduceo)
term_walko = partial(walko, rator_goal=eq, null_type=ExpressionTuple)

\end{oxtcblisting}

Listing \ref{math-expand-reduce} applies the goals to two unground logic variables, demonstrating
how miniKanren nicely covers both term expansion and reduction, as well as graph traversal
and fixed-point calculations, in a single concise framework. (The symbols
prefixed by \mintinline{python}{~_} in the output are unground logic variables.)

\begin{oxtcblisting}{minted language=python, title={\lstlistingname\ \thetcbcounter: {Simultaneous mathematical term ``expansion'' and ``reduction''.}},label type=listing, label={math-expand-reduce},nofloat}
>>> expanded_term = var()
>>> reduced_term = var()
>>> res = run(10, [expanded_term, reduced_term],
>>>           term_walko(math_reduceo, expanded_term, reduced_term))

>>> rjust = max(map(lambda x: len(str(x[0])), res))
>>> print('\n'.join((f'{str(e):>{rjust}} == {str(r)}' for e, r in res)))
                                        add(~_2291, ~_2291) == mul(2, ~_2291)
                                                   ~_2288() == ~_2288()
                              log(exp(add(~_2297, ~_2297))) == mul(2, ~_2297)
                                ~_2288(add(~_2303, ~_2303)) == ~_2288(mul(2, ~_2303))
                    log(exp(log(exp(add(~_2309, ~_2309))))) == mul(2, ~_2309)
                                             ~_2288(~_2294) == ~_2288(~_2294)
          log(exp(log(exp(log(exp(add(~_2315, ~_2315))))))) == mul(2, ~_2315)
                                           ~_2288(~_2300()) == ~_2288(~_2300())
log(exp(log(exp(log(exp(log(exp(add(~_2325, ~_2325))))))))) == mul(2, ~_2325)
                        ~_2288(~_2294, add(~_2331, ~_2331)) == ~_2288(~_2294, mul(2, ~_2331))

\end{oxtcblisting}

To further demonstrate the expressive power of miniKanren in this context, in
Listing \ref{math-constrained-expand-reduce} we show how easy it is to perform
term reduction, expansion, or both under structural constraints on the desired
terms.  Specifically, we ask for the first ten expanded/reduced term pairs where
the expanded term is a logarithm with at least one argument.

In Listing \ref{math-constrained-expand-reduce}, we use \mintinline{python}{cons}
three times to constrain the logic
variable \mintinline{python}{expanded_term} to
only \mintinline{python}{log} terms with an \mintinline{python}{add}
term as the first argument, and we'll further restrict
the \mintinline{python}{add} term to one not containing
another \mintinline{python}{add} as its first argument.

\begin{oxtcblisting}{minted language=python, title={\lstlistingname\ \thetcbcounter: {Constrained term ``expansion''" and ``reduction''.}},label type=listing, label={math-constrained-expand-reduce},nofloat}
>>> from kanren.constraints import neq
>>>
>>>
>>> log_arg = var()
>>> first_arg = var()
>>> first_arg_car, first_arg_cdr = var(), var()
>>>
>>> res = run(10, [expanded_term, reduced_term],
>>>           eq(etuple(log, log_arg), expanded_term),
>>>           eq(etuple(add, first_arg, var()), log_arg),
>>>           conso(first_arg_car, first_arg_cdr, first_arg),
>>>           neq(first_arg_car, add),
>>>           term_walko(math_reduceo, expanded_term, reduced_term))

>>> rjust = max(map(lambda x: len(str(x[0])), res))
>>> print('\n'.join((f'{str(e):>{rjust}} == {str(r)}' for e, r in res)))
>>>                             log(add((~_771 . ~_772), (~_771 . ~_772))) == log(mul(2, (~_771 . ~_772)))
>>>               log(add(log(exp(add(~_815, ~_815))), add(~_829, ~_829))) == log(add(mul(2, ~_815), mul(2, ~_829)))
>>>                           log(add((~_771 . ~_772), add(~_851, ~_851))) == log(add((~_771 . ~_772), mul(2, ~_851)))
>>>                  log(add(~_806(add(~_869, ~_869)), add(~_887, ~_887))) == log(add(~_806(mul(2, ~_869)), mul(2, ~_887)))
>>>                    log(add((~_771 . ~_772), ~_844(add(~_909, ~_909)))) == log(add((~_771 . ~_772), ~_844(mul(2, ~_909))))
>>>                           log(add(log(exp(~_788)), add(~_935, ~_935))) == log(add(~_788, mul(2, ~_935)))
>>>                 log(add((~_771 . ~_772), log(exp(add(~_945, ~_945))))) == log(add((~_771 . ~_772), mul(2, ~_945)))
>>>           log(add(~_806(~_808, add(~_967, ~_967)), add(~_985, ~_985))) == log(add(~_806(~_808, mul(2, ~_967)), mul(2, ~_985)))
>>>           log(add((~_771 . ~_772), ~_844(~_846, add(~_1007, ~_1007)))) == log(add((~_771 . ~_772), ~_844(~_846, mul(2, ~_1007))))
>>> log(add(log(exp(log(exp(add(~_1021, ~_1021))))), add(~_1035, ~_1035))) == log(add(mul(2, ~_1021), mul(2, ~_1035)))

\end{oxtcblisting}

By properly supporting \mintinline{scheme}{cons} semantics, we were able to
constructively express a non-trivial term constraint with only four simple
goals.

\subsection{The \mintinline{python}{symbolic-pymc} Package}
\label{sec:symbolic-pymc}
In order to bring together miniKanren and the popular tensor libraries Theano
and TensorFlow, our project \mintinline{python}{symbolic-pymc} provides
``meta'' type analogs for the essential tensor graph types of each ``backend'' tensor
library.  These meta types allow for more graph mutability than the ``base''
libraries themselves tend to provide.  They also allow one to use logic
variables where the base libraries wouldn't.

Basic use of \mintinline{python}{symbolic-pymc} involves either conversion
of an existing base--i.e. Theano or TensorFlow--graph into a corresponding meta graph, or
direct construction of meta graphs that are later converted (or ``reified'') to
one of the base graph types.  miniKanren goals generally work at the meta graph level,
where--for instance--one would apply goals that unify a converted base graph
with a pure meta graph containing logic variables.

While it is possible to achieve the same results with only \mintinline{python}{etuple}s,
meta graphs are a convenient form that allow developers to think and operate at the
standard Python object level.  They also provide a more direct means of graph
validation and setup, since checks can--and are--performed during meta graph construction,
whereas standard \mintinline{python}{etuple}s would not be able to perform
such operations until the resulting \mintinline{python}{etuple} is fully
constructed and evaluated.  Likewise, meta graphs are more appropriate for
specifying and obtaining derived information, like shapes and data types, for a
(sub)graph instead of miniKanren.

To demonstrate the use of \mintinline{python}{symbolic-pymc}, we consider a
simple conjugate model constructed using PyMC3 in Listing \ref{beta-binom-setup}.

\begin{oxtcblisting}{minted language=python, title={\lstlistingname\ \thetcbcounter: {A beta-binomial conjugate model in PyMC3.}},label type=listing, label={beta-binom-setup},nofloat}
import pymc3 as pm

with pm.Model() as model:
    p = pm.Beta("p", alpha=2, beta=2)
    y = pm.Binomial("y", n=totals, p=p, observed=obs_counts)

\end{oxtcblisting}

A user will generally have some data specifying the values
for \mintinline{python}{totals} and \mintinline{python}{obs_counts} and
will want to estimate the posterior distribution of \mintinline{python}{p}.
In the Bayesian world, posterior distributions are generally the object of
interest and estimation.

More specifically, we want to estimate the distribution for a rate of
success, \mintinline{python}{p}, given a total number of
events, \mintinline{python}{totals}, and observed
successes, \mintinline{python}{obs_counts}, under the assumption that the
events are binomially distributed with rate \mintinline{python}{p}.  We
let \mintinline{python}{p} take a beta distribution prior, and that
completes our Bayesian specification of a model.

Mathematically, this simple model is stated as follows:
\begin{equation}
  \label{eq:beta-binom-model}
  \begin{aligned}
    Y &\sim \operatorname{Binom}\mleft( N, p \mright)
    \\
    p &\sim \operatorname{Beta}\mleft( 2, 2 \mright)
  \end{aligned}
\end{equation}

and it has a well known closed-form posterior distribution given by
\begin{equation}
  \label{eq:beta-binom-post}
  \left( p \mid Y=y \right) \sim \operatorname{Beta}\mleft( 2 + y, 2 + N - y \mright)
\end{equation}

Instead of wasting resources estimating the posterior numerically
(e.g. using \mintinline{python}{pm.sample(model)} to run a Markov Chain
Monte Carlo sampler), we can simply extract the underlying Theano graph
from \mintinline{python}{model} and apply a relation that represents the
underlying conjugacy and use the resulting posterior.

The general rule implied by this situation is
\begin{equation}
  \label{eq:beta-binom-rule}
  \begin{aligned}
    \left( \text{beta-binomial-conjugate} \right) &\quad
    \frac{
      Y \sim \operatorname{Binom}\mleft( N, p \mright), \quad
      p \sim \operatorname{Beta}\mleft( \alpha, \beta \mright)
    }{
      \left( p \mid Y=y \right) \sim \operatorname{Beta}\mleft( \alpha + y, \beta + N - y \mright)
    }
  \end{aligned}
\end{equation}

Listing \ref{beta-binom-convert} converts the PyMC3 model object, \mintinline{python}{model},
into a standard Theano graph that represents the relationship between random variables
in the model.

\begin{oxtcblisting}{minted language=python, title={\lstlistingname\ \thetcbcounter: {Converting a PyMC3 model into a Theano graph of the model's sample-space.}},label type=listing, label={beta-binom-convert},nofloat}
from symbolic_pymc.theano.pymc3 import model_graph
from symbolic_pymc.theano.utils import canonicalize


# Convert the PyMC3 graph into a symbolic-pymc graph
fgraph = model_graph(model)

# Perform a set of standard algebraic simplifications using Theano
fgraph = canonicalize(fgraph, in_place=False)

\end{oxtcblisting}

Listing \ref{beta-binom-goal} uses miniKanren to construct a
goal, \mintinline{python}{betabin_conjugateo}, that matches terms taking the
form of Equation \eqref{eq:beta-binom-model} in the first argument and the
resulting posterior of Equation \eqref{eq:beta-binom-post} in the second argument.
It makes use of both meta objects and \mintinline{python}{etuple}s.

\begin{oxtcblisting}{minted language=python, title={\lstlistingname\ \thetcbcounter: {A miniKanren goal that implements the beta-binomial conjugate rule in \eqref{eq:beta-binom-rule}.}},label type=listing, label={beta-binom-goal},nofloat}
def betabin_conjugateo(x, y):
    """Replace an observed Beta-Binomial model with an unobserved posterior Beta-Binomial model."""
    obs_lv = var()

    beta_size, beta_rng, beta_name_lv = var(), var(), var()
    alpha_lv, beta_lv = var(), var()
    # We use meta objects directly to construct the "pattern" we want to match
    beta_rv_lv = mt.BetaRV(alpha_lv, beta_lv, size=beta_size, rng=beta_rng, name=beta_name_lv)

    binom_size, binom_rng, binom_name_lv = var(), var(), var()
    N_lv = var()
    binom_lv = mt.BinomialRV(N_lv, beta_rv_lv, size=binom_size, rng=binom_rng, name=binom_name_lv)

    # Here we use etuples for the output terms
    obs_sum = etuple(mt.sum, obs_lv)
    alpha_new = etuple(mt.add, alpha_lv, obs_sum)
    beta_new = etuple(mt.add, beta_lv, etuple(mt.sub, etuple(mt.sum, N_lv), obs_sum))

    beta_post_rv_lv = etuple(
        mt.BetaRV, alpha_new, beta_new, beta_size, beta_rng, name=etuple(add, beta_name_lv, "_post")
    )
    binom_new_lv = etuple(
        mt.BinomialRV,
        N_lv,
        beta_post_rv_lv,
        binom_size,
        binom_rng,
        name=etuple(add, binom_name_lv, "_post"),
    )

    return lall(eq(x, mt.observed(obs_lv, binom_lv)), eq(y, binom_new_lv))


\end{oxtcblisting}

Finally, Listing \ref{beta-binom-run} shows how the goal can be applied to the model's graph
and how a new Theano graph and PyMC3 model is constructed from the output.

\begin{oxtcblisting}{minted language=python, title={\lstlistingname\ \thetcbcounter: {Running the beta-binomial conjugate goal and creating a PyMC3 model for the results.}},label type=listing, label={beta-binom-run},nofloat}
from symbolic_pymc.theano.pymc3 import graph_model


q = var()
res = run(1, q, betabin_conjugateo(fgraph.outputs[0], q))

expr_graph = res[0].eval_obj
fgraph_conj = expr_graph.reify()

# Convert the Theano graph into a PyMC3 model
model_conjugated = graph_model(fgraph_conj)

\end{oxtcblisting}

\citet{WillardTourSymbolicPyMC2020} gives a more thorough walk-through
of \mintinline{python}{symbolic-pymc} and miniKanren--using TensorFlow
graphs.

\section{Discussion}
\label{sec:orgd60a5fa}

Looking forward, the ``functions grimoire'' project of
\citet{Johanssonfungrim2020}, Fungrim, is a great example of community-sourced and
programmatically encoded domain knowledge in mathematics.  It serves as a prime
example of an independently developed, high-level knowledge encoding effort from
which relations could be sourced and used by miniKanren in Python.  In
\citet{ByrdwebyrdmediKanren2020} miniKanren is used to reason over an external
database of medical knowledge, so the a precedent for this type of work has
already been set.

One currently unexplored area involves interactions between miniKanren and
symbolic algebra libraries.  Although a lot of symbolic algebra is possible
using miniKanren alone, we don't necessarily expect it to replace symbolic
algebra libraries any time soon.  Luckily, when our statistical modeling goals require
advanced symbolic algebra functionality provided by existing libraries, like
SymPy \citep{sympydevelopmentteam_sympy_2014},
they can be used directly from within miniKanren.
For example, one could use SymPy to perform a Laplace transform--or its
inverse--within an implementation of a normal scale mixture
\citep{BhadraGlobalLocalMixtures2020} relation between random variables, since
the underlying mathematical relation is functionally described by Laplace
transforms in both directions.

A combination of miniKanren and computer algebra could also be used to realize
elements of the computer algebra and interactive theorem proving synthesis
described in \citet{KaliszykCertifiedcomputeralgebra2007}.

Perhaps another concrete example of how a symbolic algebra library could be leveraged
by miniKanren is given by the system described in
\citet{Waliahighlevelinferencealgorithms2018}.  Here, miniKanren could be used
to implement the typing rules, and the \texttt{integrate} steps could be outsourced to
SymPy.  By targeting standard NumPy \citep{NumpyDevelopersNumpy2017} output,
the results could be used by any number of systems, like JAX, that provide
vectorization, JIT compilation, etc.

Ultimately, we've described the construction of a strictly Python version of the
system in \citet{Waliahighlevelinferencealgorithms2018} that makes use of
multiple popular, actively developed projects specializing in their respective
domains (e.g. symbolic integration, vectorization, JIT compilation).  The original
implementation uses a complex pipeline
\citep[Figure 1]{Waliahighlevelinferencealgorithms2018} that operates across
multiple independent systems, one of which is Hakaru
\citep{NarayananProbabilisticinferenceprogram2016}.  Hakaru is an entirely
independent system and programming language for probabilistic programming.  In
contrast, the system we describe is simply an orchestration of existing Python
libraries driven by miniKanren, and it can make use of whichever tensor libraries,
compilation frameworks, and PPLs are most suitable.

Regarding miniKanren itself, consider the following idiom:
\begin{oxtcblisting}{minted language=scheme, nofloat}
(conde ((== lhs match-form-1)
        (== rhs replace-form-1))
       ((== lhs match-form-2)
        (== rhs replace-form-2))
       ...)

\end{oxtcblisting}
\mintinline{scheme}{conde}s like this are natural for encoding identities
of the form \(\text{match-form-1} = \text{replace-form-1}\) that are applied to
the terms \mintinline{scheme}{lhs} and \mintinline{scheme}{rhs}.  They
also appear in implementations of relational interpreters where they encode the
supported forms of a target language (e.g. variable assignment, conditionals,
etc.)

These \mintinline{scheme}{conde} idioms comprise a large portion of the
miniKanren work implied here, and their size could grow very quickly over time.
This leads to performance questions that are possibly answered by work on guided
search \citep{SwordsGuidedSearchminiKanren,ZhangNeuralGuidedConstraint2018}
and discerning \mintinline{scheme}{conde} branch selection
\citep{BoskinSurprisinglyCompetitiveConditional2018}.

There is also little reason to think that a
single \mintinline{scheme}{conde} will--or even \emph{should}--encode most of a
system's implemented identities, so a means of compiling goals--say--for the
purposes of merging branches might be worth considering, as well.

With equational identities nicely encoded by \mintinline{scheme}{conde}
forms, there's also the possibility that the rewrite-completion algorithms
mentioned in Section \ref{sec:want-to-do} could be applied automatically.  When a
complete and reduced rewrite system can be generated from
a \mintinline{scheme}{conde}, it would be interesting to know whether or not
the resulting system improves the general performance of miniKanren.

Also, is it possible that some cases of non-terminating goal orderings could be
avoided by completion? Likewise, could the results of completion be used to
produce a new, equivalent \mintinline{scheme}{conde} that results in fewer
goal evaluations and/or failed branches?

Alternatively, is it possible that miniKanren could be utilized by a
completion algorithm itself so that it produces potentially relevant results
when it otherwise wouldn't terminate (e.g. via infinite goal streams)?
What are the advantages of performing completion in a relational fashion, and
what unique elements can miniKanren provide to that situation (e.g. easier
implementation of experimental completion algorithms)?

Proving rewrite termination and completion itself can involve SAT problems
\citep{EndrullisMatrixinterpretationsproving2008,KleinMaximalcompletion2011};
can miniKanren's constraint capabilities--among other things--be applied in this
area?

Our Python implementation of miniKanren comes with experimental
support for associative and commutative (AC) relations.
We've found utility in assigning these two properties to existing operators from
other libraries (e.g. addition operators in Theano and TensorFlow) as a means
of adding flexibility to the exact representation of graphs.  This is especially
important in instances where graph normalization isn't entirely consistent or
available via the targeted graph backend (e.g. TensorFlow).

The process of implementing these AC relations has opened a few questions that
cannot be properly treated here.  Questions such as "How can operators with
arbitrary--but known and fixed--arities be efficiently supported?" and "How can
we overcome some of the goal ordering issues that arise due to commutativity?".

In \citep{WillardPullRequest272020}, we address the latter question with a
``groundedness''-based term reordering goal.  This reordering is performed on
the \mintinline{scheme}{cdr} sub-terms as a relation is applied between term
graphs, since the order in which a relation is applied to corresponding
sub-terms is--generally--immaterial.  In other words, when walking a
goal \mintinline{scheme}{relo} between the
lists \mintinline{scheme}{'(a b)} and \mintinline{scheme}{'(c 2)}, for fresh
variables \mintinline{scheme}{a}, \mintinline{scheme}{b}, and \mintinline{scheme}{c}, the
goal can be applied in any order, e.g. \mintinline{scheme}{(relo a c)}
then \mintinline{scheme}{(relo b 2)}, or \mintinline{scheme}{(relo b 2)}
then \mintinline{scheme}{(relo a c)}.  When--for
instance--\mintinline{scheme}{(relo a b)} diverges because both arguments
are fresh, while \mintinline{scheme}{(relo b 2)} fails, a walk that performs
the former ordering will diverge, while one that does the latter will fail.  The
``groundedness'' ordering goal simply reorders the corresponding pairs according
to how grounded they are to arrive at the non-diverging order of application.

Finally, we would like to point out the potential for an exciting ``feedback loop'': as
statistical modeling improves the processing of miniKanren
\citep{ZhangNeuralGuidedConstraint2018}, miniKanren can also improve the
process of statistical modeling.

\begin{acks}
The author would like to thank Jason Hemann and William Byrd for their
invaluable input and inspiring work.

\end{acks}

\bibliographystyle{plainnat}
\bibliography{ICFP2020}
\end{document}